\documentclass[aps,pre,showpacs,twocolumn,floatfix,amssymb,superscriptaddress]{revtex4-1}

\usepackage{graphicx}
\usepackage[intlimits]{amsmath}
\usepackage{color}

\newcommand{\modified}[1]{#1}

\begin{document}

\title{Suppression of interference in quantum Hall Mach-Zehnder geometry by upstream neutral modes}

\author{Moshe Goldstein}
\affiliation{Raymond and Beverly Sackler School of Physics and Astronomy, Tel Aviv University, Tel Aviv 6997801, Israel}

\author{Yuval Gefen}
\affiliation{Department of Condensed Matter Physics, The Weizmann Institute of Science, Rehovot 76100, Israel}

\begin{abstract}
Mach-Zehnder interferometry has been proposed as a probe for detecting the statistics of anyonic quasiparticles in fractional quantum Hall (FQH) states. Here we focus on interferometers made of multimode edge states 
with upstream modes. We find that the interference visibility is suppressed due to downstream-upstream mode entanglement; the latter serves as a ``which path'' detector to the downstream interfering trajectories. Our analysis tackles a concrete realization of filling factor $\nu=2/3$, but its applicability goes beyond that specific case, and encompasses the recent observation of ubiquitous emergence of upstream neutral modes in FQH states. The latter, according to our analysis, goes hand in hand with the failure to observe Mach-Zehnder anyonic interference in fractional states.  We point out how charge-neutral mode disentanglement will resuscitate the interference signal.
\end{abstract}


\maketitle

\emph{Introduction.---}
One of the most striking implications of the theory of the fractional quantum Hall (FQH) effect is the nature of the elementary excitations (``anyons''), featuring fractional charge and fractional statistics.
The latter concerns the change in the state of the system upon braiding the anyons: it is multiplied by a phase in the abelian case, and by a unitary operator in the non-abelian case, thus raising the prospect of topologically-protected quantum computation \cite{nayak08}.
Experimental evidence for fractional charge dates back almost two decades ago \cite{goldman95,depicciotto97,saminadayar97}.
Notwithstanding intriguing results \cite{camino0507,willett0910,ofek10}, 
fractional statistics has remained elusive to date.
A natural probe of statistics is through its effect on the Aharonov-Bohm (AB) interferometry of anyons moving along the gapless edges of a FQH system. Mach-Zehnder interferometers \cite{ji03,neder06,neder07b,litvin07,roulleau07,roulleau08,litvin08} would have been efficient tools to observe anionic interferometry  as they avoid interference-masking Coulomb effects \cite{rosenow07,halperin11} and are robust against further quasiparticle fluctuations in the bulk \cite{law06,feldman06,feldman07}.

The interfering paths of a Mach-Zehnder interferometer (MZI) rely on the chiral edge modes of the FQH geometries. Multimode edges \cite{macdonald90,johnson91,meir94} present one with an interesting twist: the possibility of upstream moving modes \modified{(i.e., modes moving against the ``downstream'' direction set by the magnetic field)}. These may be neutral \modified{\cite{kane94,kane95,cano14}}.
Such modes have been experimentally detected through the generation of upstream charge noise \cite{bid10,dolev11,gross12,gurman12} and 
thermometry \cite{venkatachalam12}.
Recent measurements \cite{inoue14} have surprisingly found that, unlike earlier predictions, upstream neutral modes are not restricted to ``hole-like'' states (e.g., $1/2<\nu<1$), but rather show up in virtually \emph{all} FQH states, including simple ``electron-like'' Laughlin states (such as $\nu=1/3$) \cite{fn:upstream}. How do these upstream moving modes affect the expected anyonic interference? 

Here we show that their effect on the anyonic interference visibility is detrimental. 
The present analysis, employing the example of a $\nu=2/3$ FQH system \cite{kane94,kane95},  entails a single upstream-propagating neutral mode along the edge of a MZI. We note that our picture holds for more complex edge profiles than the minimal ones dictated by the bulk properties \cite{wang13}, and for other setups which support upstream moving modes, be them neutral or charged. 
We find that the  suppression of the interference visibility is due to downstream-upstream mode entanglement: the upstream (e.g., neutral)  mode serves as a ``which path'' detector of the downstream (charged) quasiparticle trajectory. We analyze the dependence of the entanglement on the system's characteristics (geometry, temperature, voltage bias), and discuss the parameter regime where the charge and neutral degrees of freedom may be disentangled, leading to recuperation of the interference signal.

From a broader perspective, here we explore how these two experimental facts, namely the 
undetectability of anyonic AB fringes in MZI (unlike the high visibility of interference in the integer QH regime \cite{ji03,neder06,neder07b,litvin07,roulleau07,roulleau08,litvin08}), and the ubiquity of upstream neutral modes in FQH systems, are related. We show that the latter leads to an exponential suppression of the former. This is the case even with an equal-arms-length MZI, where thermal averaging is avoided in standard scenarios.

\begin{figure}
\includegraphics[width=8cm,height=!]{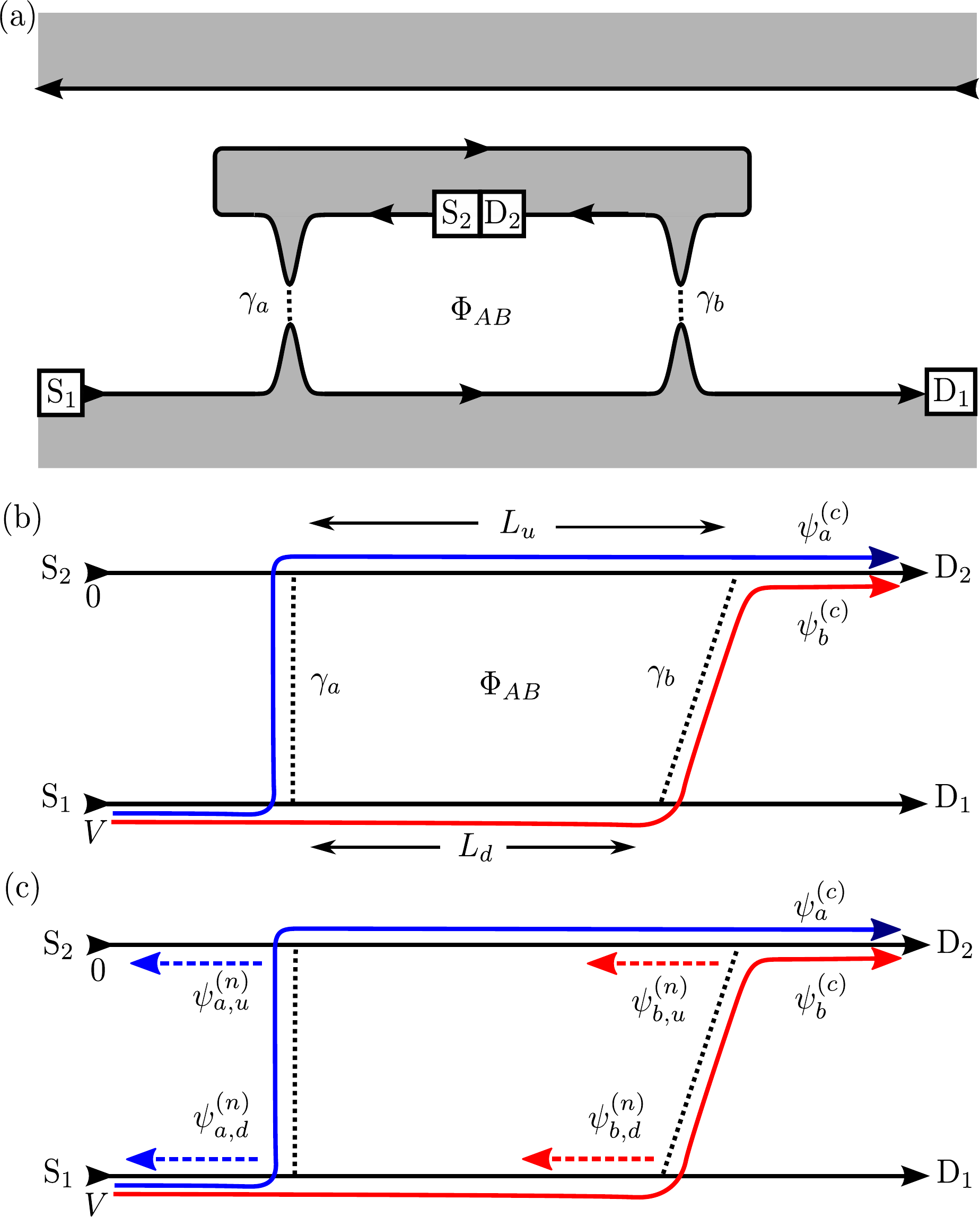}
\caption{\label{fig:interference}
(Color online) QH MZI: (a) Experimental setup of a MZI. Shown are the downstream chiral edge modes, sources S$_1$ and S$_2$ (between which a bias voltage $V$ is applied), drains D$_1$ and D$_2$, and the tunneling bridges a and b. A schematic equivalent geometry is depicted in (b,c), whose lower and upper edges correspond to the lower and island edges in (a), respectively. Also shown in (b,c) is the AB interference of partial wavelets approaching D$_2$ (to order $\gamma_a \gamma_b$).
(b) $\nu=1/3$ Laughlin state. Two partial waves of a charge-$e/3$ quasiparticle tunnel through QPCs $a$ and $b$, respectively. The interference visibility is proportional to the overlap of $\psi^{(c)}_a$ and $\psi^{(c)}_b$.
(c) $\nu=2/3$ KFP 
\cite{kane94,kane95} state. Each charge tunneling event is accompanied by the emission of two neutral jets (dashed arrows). The AB signal at terminal D$_2$ is then proportional to the product of the overlaps of ($\psi^{(c)}_a$, $\psi^{(c)}_b$), ($\psi^{(n)}_{a,u}$, $\psi^{(n)}_{b,u}$), and ($\psi^{(n)}_{a,d}$, $\psi^{(n)}_{b,d}$).
The neutral jets $\psi^{(n)}_{a,u}$ and $\psi^{(n)}_{b,u}$ (and similarly $\psi^{(n)}_{a,d}$ and $\psi^{(n)}_{b,d}$) are emitted at different times and at different spatial points (even for a symmetric interferometer, $L_d=L_u$), leading to interference suppression.}
\end{figure}

\emph{The main idea} is depicted in Fig.~\ref{fig:interference}. In general, the manifestation of interference of two paths (``coherency'') requires that the difference between the propagation time along each one of them is smaller than the inverse of the energy window $\Delta \varepsilon$ over which one is averaging (coherence time), e.g., the thermal spread $\Delta \varepsilon \sim k_B T$.
When only a single downstream charge mode with velocity $v_c$ is present [Fig.~\ref{fig:interference}(b)], it follows that observing interference requires
$\hbar / \Delta \varepsilon > |L_d-L_u|/v_c$.
In a real-space picture, this means that there should be a significant overlap between the two interfering charge wavelets, i.e., that their spatial separation, $|L_d-L_u|$, is smaller than the wavelet size $\hbar v_c / \Delta \varepsilon$.
From both perspectives, for a symmetric interferometer  ($L_d=L_u$), energy averaging due to the finite width ($\sim k_B T$) of the incoming wavepackets does not undermine the interference.

In the presence of an upstream neutral mode (with velocity $v_n$) [Fig.~\ref{fig:interference}(c)], tunneling through the bridges (or quantum point contacts, QPCs) a or b involves the excitation of neutral \modified{wavepackets} or ``jets''.
\modified{Thus, the interfering amplitudes at D$_2$ can be schematically written as
$|\Psi \rangle_{D2} =
| \psi^{(c)}_a \rangle \otimes | \psi^{(n)}_{a,u} \rangle \otimes | \psi^{(n)}_{a,d} \rangle
+ e^{i \theta} | \psi^{(c)}_b \rangle \otimes | \psi^{(n)}_{b,u} \rangle \otimes | \psi^{(n)}_{b,d} \rangle$,
with $\theta$ the sum of the AB and the orbital phase differences between the two paths.
For interference to show up, there must be nonvanishing overlap between corresponding pairs of neutral wavepackets,
$\langle \psi^{(n)}_{a,u} | \psi^{(n)}_{b,u} \rangle$ and
$\langle \psi^{(n)}_{a,d} | \psi^{(n)}_{b,d} \rangle$.}
Note that $\psi^{(n)}_{a,d}$ and $\psi^{(n)}_{b,d}$, for example, are separated in space and time, as it takes the charge mode some time to propagate from QPC a to QPC b. Their separation in time is then $L_d/v_c+L_d/v_n$ (it is $L_d/v_c+L_u/v_n$ for the neutral jets on the upper arm). The overlap between neutral jets decreases with the overall size of the MZI, and cannot be compensated by resorting to a symmetric setup,
\modified{but only by reducing the temperature and voltage. Due to the expected smallness of the neutral mode velocity (stiffness), these restrictions can be quite severe in practice.}

\emph{Model.---}
To substantiate this idea we focus on the Kane-Fisher-Polchinski (KFP) \cite{kane94,kane95} model of the $\nu=2/3$ FQH state.
The latter can be thought of as a $\nu=1/3$ state of holes on top of a $\nu=1$ state of electrons \cite{macdonald90,johnson91,meir94}. Correspondingly, the edge theory consists of a downstream $\nu=1$ mode $\phi_1$ and an upstream $\nu=1/3$ mode $\phi_2$.
The inclusion of disorder typically leads to equilibration between the edge modes, and results in a universal fixed point (with nonuniversal mode velocities). There the charge and neutral sectors, given by the linear combinations $\phi_c = \sqrt{3/2}(\phi_1 + \phi_2)$ and $\phi_n = -(\phi_1+3\phi_2)/\sqrt{2}$, respectively, are decoupled.

The operators
$\psi_{n_1,n_2}^\dagger =e^{i n_1 \phi_1 + i n_2 \phi_2} = e^{i \kappa_c \phi_c + i \kappa_n \phi_n}$, with integer $n_1$ and $n_2$, correspond to the creation of a quasiparticle of charge $Q=e(n_1-n_2/3)$.
The scaling dimension of an inter-edge tunneling process involving such a quasiparticle at the KFP 
fixed point is $(\kappa_c^2 + \kappa_n^2)/2$.
The statistical (braiding) phase of two such quasiparticles with quantum numbers $n_{1,2}$ and $m_{1,2}$, respectively, is $\pi (n_1 m_1 - n_2 m_2/3)$.
For an almost-open QPC it is sufficient to address the most relevant \modified{(in the renormalization group sense)} tunneling operators, involving three types of quasiparticles (cf.\ Table~\ref{tbl:qps}).
\modified{Of these, only $\psi^\dagger_{0,-1}$ includes just the innermost mode $\phi_2$, hence it should have the largest bare tunneling amplitude and give the dominant contribution.}
Similarly, for an almost-closed QPC, one needs to consider the most relevant processes in which \emph{integer} charge (electrons) is transferred. Of these two (see Table~\ref{tbl:qps}), the most important is $\psi^\dagger_{1,0}$, as it depends only on the outermost mode $\phi_1$. In either case the most important tunneling operator involves both charge and neutral modes,
hence tunneling of charge excites the neutral modes on both sides of the QPC.
Below we first consider tunneling processes involving only $\psi^\dagger_\textrm{max} = \psi^\dagger_{0,-1}$ (quasiparticle) or $\psi^\dagger_{1,0}$ (electron), respectively. We later discuss what happens when all the most relevant tunneling operators listed in Table~\ref{tbl:qps} are present, and show that this 
does not modify our results in an essential way.

\begin{table}
\begin{ruledtabular}
\begin{tabular}{l|cccccc}
  & $n_1$ & $n_2$ & $Q/e$ & $\kappa_c$   & $\kappa_n$    & Scaling dimension \\ \hline
  Fractional-charge& 0    & -1     & 1/3   & $1/\sqrt{6}$ & $1/\sqrt{2}$  & 1/3 \\ 
  quasiparticles & 1     & 2     & 1/3   & $1/\sqrt{6}$ & $-1/\sqrt{2}$ & 1/3 \\
  & 1     & 1     & 2/3   & $\sqrt{2/3}$ & 0             & 1/3 \\ \hline
  Integer charge& 1     & 0     & 1     & $\sqrt{3/2}$ & $1/\sqrt{2}$  & 1   \\
  (electron) & 2     & 3     & 1     & $\sqrt{3/2}$ & $-1/\sqrt{2}$ & 1   \\
\end{tabular}
\end{ruledtabular}
\caption{\label{tbl:qps}
Quantum numbers of the most relevant quasiparticle and electron creation operators $\psi_{n_1,n_2}^\dagger =e^{i n_1 \phi_1 + i n_2 \phi_2} = e^{i \kappa_c \phi_c + i \kappa_n \phi_n}$ \cite{kane94,kane95,viola12,kamenev15}.}
\end{table}

Under the above assumptions, the action of the MZI accounts for the two constituent edges $\ell=d,u$ (in units where $\hbar=k_B=1$), 
\begin{align}
  \mathcal{S}_{0} = \frac{1}{4\pi} \sum_{\ell=d,u} \int \mathrm{d}t \int \mathrm{d}x &
  \partial_x \phi_{\ell}^{c} \left(-\partial_t \phi_{\ell}^c - v_c \partial_x \phi_{\ell}^c \right)
  \\ \nonumber &
  + \partial_x \phi_{\ell}^n \left(\partial_t \phi_{\ell}^n -v_n \partial_x \phi_{\ell}^n \right),
\end{align}
and the tunneling bridges 
at the two QPCs, $i=a,b$,
\begin{align}
  \mathcal{S}_\mathrm{QPCs} =
  -\sum_{i=a,b} & \int \mathrm{d}t
  \gamma_i \eta_i e^{i 2\pi \Phi_{AB}/\Phi_0 \delta{i,a} + i e^* V (t-x_{i,d}/v_c)}
  \nonumber \times \\ &
  \psi^\dagger_{d,\mathrm{max}}(x_{i,d},t) \psi_{u,\mathrm{max}}(x_{i,u},t)
  + \mathrm{c.c.}
\end{align}
Here $\gamma_i$ are the tunneling amplitudes, $V$ is the source-drain voltage applied between S$_1$ and S$_2$, $\Phi_{AB}$ is the AB flux enclosed inside the interference loop ($\Phi_0$ is the flux quantum),
and
$\psi_{\ell,\mathrm{max}}^\dagger = e^{i\kappa_c \phi^c_\ell + i \kappa_n \phi^n_\ell}$,
with
$\kappa_n = 1/\sqrt{2}$, $\kappa_c = 1/\sqrt{6}$ ($\kappa_c=\sqrt{3/2}$), and $e^*=e/3$ ($e^*=e$) for an almost open (closed) QPC.
The coordinates of the QPCs on the two edges are $x_{a,d} = x_{a,u} = 0$, $x_{b,d} = L_d$, and $x_{b,u} = L_u$.
Finally, $\eta_i$ are Klein factors \cite{law06}. These
are essential in the case of quasiparticle tunneling (almost open QPCs):
the Klein factors account for the fact that fractional excitations cannot leave the QH edge into the drain; rather they 
wait there until an integer charge is accumulated, and are then absorbed by the drain. The quasiparticles that wait at D$_2$ contribute [through the braiding phase; cf. Fig.~\ref{fig:interference}(a)] to the interference phase of the quasiparticles that follow.
For now we are considering only a single tunneling process;
the Klein factor may therefore be taken to be the same
as in the $\nu=1/3$ case \cite{law06,fn:disorder}.

\begin{table}
\begin{ruledtabular}
\begin{tabular}{c|l}
  Regime            & Visibility \\ \hline
  $|e^* V|, T \ll 1/t_L$ & $\sim 1$ \\
  $|e^* V|, 1/t_L \ll T$ & $\sim \exp[- 2 \pi (e/e^*) \kappa_\text{min}^2 T t_L]$\\
  $1/t_L \ll T \ll |e^* V|$ & $\sim |e^* V/T|^{-2\kappa_\text{min}^2} \exp[- 2 \pi (e/e^*) \kappa_\text{min}^2 T t_L]$ \\
  $T \ll 1/t_L \ll |e^* V|$ & $\sim |e^* V t_L|^{-2\kappa_\text{min}^2} \times$ \\
  & $\phantom{\sim}\left[1+ C_0 |e^* V t_L|^{-2|\kappa_c^2-\kappa_s^2|}\cos(e^* V t_L - \phi_0) \right] $ \\
\end{tabular}
\end{ruledtabular}
\caption{\label{tbl:visibility}
Interference visibility (ratio between the amplitude of the $\Phi_{AB}$-periodic contribution to the current and the $\Phi_{AB}$-independent contribution to the current) of a symmetric MZI in the $\nu=2/3$ KFP 
state,
for either quasiparticle tunneling (almost open QPCs, $e^*=e/3$, $\kappa_c=1/\sqrt{6}$, $\kappa_n=1/\sqrt{2}$),
or electron tunneling (almost closed QPCs, $e^*=e$, $\kappa_c=\sqrt{3/2}$, $\kappa_n=1/\sqrt{2}$).
$t_L=L/v_c+L/v_n$ is the delay time, $\kappa_\text{min}=\min(\kappa_c,\kappa_n)$,
$C_0 = 2 (e/e^*) [\Gamma(2\kappa_\text{min}^2)]^2 / [\Gamma(2\kappa_c^2)\Gamma(2\kappa_n^2)]$,
and
$\phi_0 = \pi (\kappa_c^2+\kappa_n^2-1)$.
}
\end{table}

\emph{Analysis and results.---}
To simplify the presentation, we henceforth concentrate on the limit of almost open QPCs; the results for almost closed QPCs are similar, and are discussed in the Supplemental Material \cite{sm}.
Working to lowest order in the tunneling amplitudes, $\gamma_i$, one finds for the current at D$_2$ \cite{law06},
\begin{equation} \label{eqn:id2}
  \frac{1}{I_{\mathrm{D}_2}(V,T)} =
  \frac{1}{3} \left[ \frac{1}{I_{0}(V,T)} + \frac{1}{I_{1}(V,T)} + \frac{1}{I_{2}(V,T)} \right],
\end{equation}
with $I_{k}(V,T)$ being the average current at D$_2$, given that $k=0,1,2$ (mod 3) anyons have already reached D$_2$,
\begin{align} \label{eqn:ik}
  I_{k}(V,T) = &
  \left(\left| \gamma_a \right|^2 + \left| \gamma_b \right|^2 \right) j(0,0;V,T)
  +
  \\ &
  \gamma_a \gamma_b^* e^{2 \pi i (e^*/e) \Phi_{AB}/\Phi_0 - 2\pi i k / 3} j(L_d,L_u;V,T) + \text{c.c.}
  \nonumber
\end{align}
Here
\begin{align} \label{eqn:j}
  j(y_d, y_u;V,T) = &
  e^* \left[ 1 - e^{- e^* V/T } \right]
  \times \\ \nonumber &
  \int_{-\infty}^{\infty} \mathrm{d}t e^{i e^* V (t - y_d/v_c)} F(y_d,y_u,t;T),
\end{align}
depends on the correlation function of the tunneling operators at $\gamma_a=\gamma_b=0$,
\begin{align} \label{eqn:f}
  & F(y_d,y_u,t;T) \nonumber \\
  & = \left\langle
  \psi^\dagger_{u,\mathrm{max}}(y_u,t) \psi_{d,\mathrm{max}}(y_d,t)
  \psi^\dagger_{d,\mathrm{max}}(0,0) \psi_{u,\mathrm{max}}(0,0) 
  \right\rangle_{\gamma_i=0}
  \nonumber \\
  & =
  \left\{
  \frac{\pi T \tau}{i \sinh \left[ \pi T ( t^- - y_d/v_c) \right]}
  \right\}^{\kappa_c^2}
  \negthickspace
  \left\{
  \frac{\pi T \tau}{i \sinh \left[ \pi T ( t^- - y_u/v_c) \right]}
  \right\}^{\kappa_c^2}
  \times \nonumber \\ &
  \hphantom{ = }
  \left\{
  \frac{\pi T \tau}{i \sinh \left[ \pi T ( t^- + y_d/v_n) \right]}
  \right\}^{\kappa_n^2}
  \negthickspace
  \left\{
  \frac{\pi T \tau}{i \sinh \left[ \pi T ( t^- + y_u/v_n) \right]}
  \right\}^{\kappa_n^2},
\end{align}
where $t^- = t - i\delta$ \modified{($\delta \to 0^+$) and $\tau$ is the inverse ultraviolet cutoff, of the order of the bulk gap}.
Should the neutral modes be absent, $j(y_d,y_u,t;T)$ would depend only on $y_d-y_u$ [as can be seen from Eq.~(\ref{eqn:j}) upon a shift of the integration variable $t$ by $y_d /v_c$]; in particular, for a symmetric interferometer, $L_d=L_u=L$ and $|\gamma_a|=|\gamma_b|$, the amplitude of the flux-dependent contribution to Eq.~(\ref{eqn:ik}) has the same magnitude as the flux-independent term (barring additional dephasing mechanisms beyond our model), and the visibility can reach unity even at finite $V$ and $T$.
Taking the neutral \modified{modes} into account, $j(y_d,y_u,t;T)$ depends on both $(y_d-y_u)/v_c$ and $y_d/v_c+y_{d/u}/v_n$ . It follows that the visibility is suppressed at finite $e^* V$ and $T$ even in the symmetric case, once either energy scale becomes \modified{larger} than the inverse of the delay time
\modified{(between neutral modes emitted when charge tunnels at either QPC)}
$t_L \equiv L/v_c + L/v_n$ \cite{fn:disorder,fn:three}.
A summary of the results in this case is given in Table~\ref{tbl:visibility} and Fig.~\ref{fig:visibility}.
Detailed analysis of the general case is presented in the Supplemental Material \cite{sm}.

\begin{figure}
\includegraphics[width=8cm,height=!]{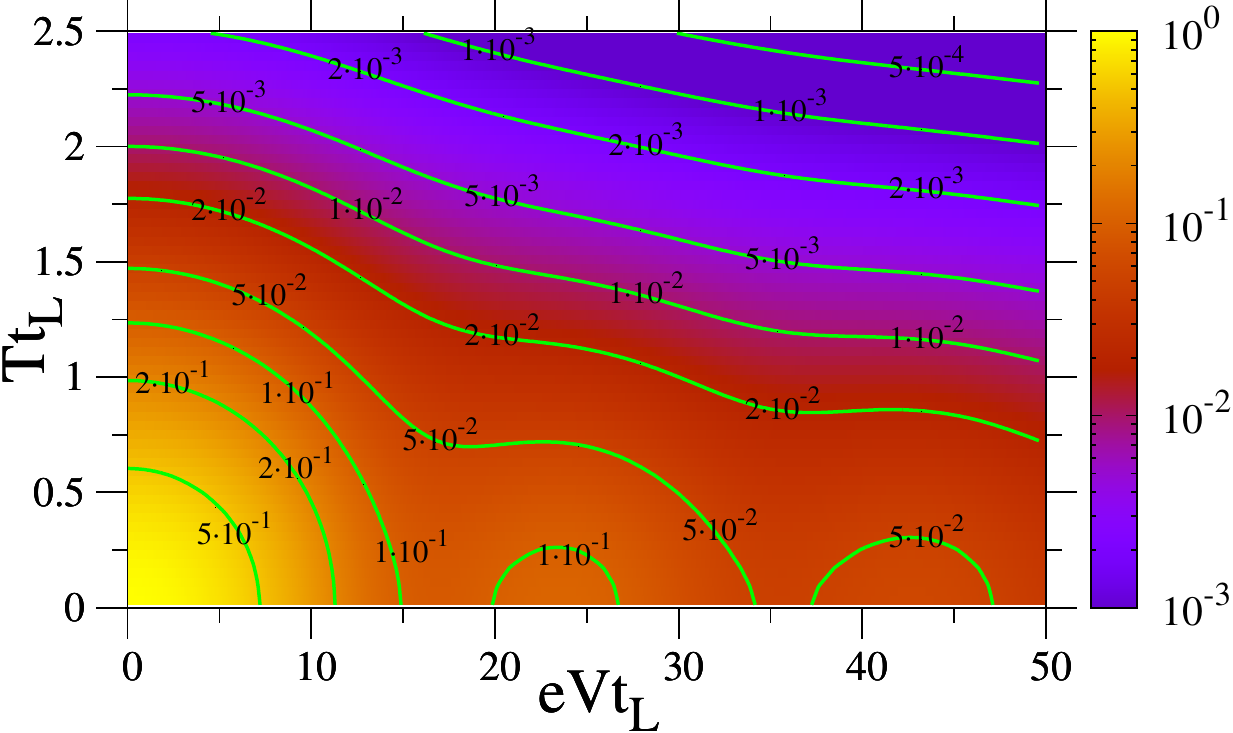}
\caption{\label{fig:visibility} (Color online)
Color map of the interference visibility (ratio between the amplitude of the $\Phi_{AB}$-periodic contribution to the current and the $\Phi_{AB}$-independent contribution to the current) of a symmetric MZI in the $\nu=2/3$ KFP 
(quasiparticle tunneling) as function of temperature and voltage.
\modified{The delay time $t_L=L/v_c+L/v_n$ corresponds (for $L \approx 5$~$\mu$m, $v_n \approx 10^5$~cm/sec) to $0.1$~$\mu$V or $1.5$~mK.
Note the lobe structure in the low-$T$ and high-$V$ regime; see Table~\ref{tbl:visibility} (last line) and the Supplemental Material~\cite{sm}.}
}
\end{figure}

\emph{Accounting for all the most relevant quasiparticle tunneling operators.---}
Up to this point we have restricted ourselves to the case where only the charge-$e/3$ tunneling operator involving the inner-most edge mode (cf.~Table~\ref{tbl:qps}) is considered, since it is expected to have the largest amplitude in the weak tunneling regime. The tunneling quasiparticle fractionalizes into the charge and the neutral modes, 
which results in the suppression of the interference signal.
Including the other $e/3$ most-relevant operator (again with charge-neutral fractionalization) will not modify the results qualitatively.
On the other hand, the charge-$2e/3$ most relevant operator creates \emph{no neutral excitations}, hence it \emph{may} contribute to the interference.
However, let us first note that the bare amplitude of the $2e/3$ tunneling operator should be much smaller than that of the $e/3$ ones. We then recall that in the fractional case the phase shift of the AB signal depends on the number of quasiparticles which have tunneled earlier and got trapped at D$_2$ [Fig.~\ref{fig:interference}(a)]. It follows that the tunneling of charge-$e/3$ quasiparticles shifts the interference signal of the charge-$2e/3$ quasiparticles erratically by $\pm 2\pi/3$, thus undermining the visibility.
To see that, let us focus on the regime $|eV| \gg T$, where only tunneling from high to low voltage 
is allowed.
We will further assume that $\max(L_d,L_u)/v_n \gg \max[1/T, 1/|eV|]$, so that processes involving neutral excitations do not contribute to the interference at all. We may then describe the dynamics employing a master equation. It has three states, corresponding to charge $k e/3$ accumulated at D$_2$, modulo the electron charge (i.e., $k=0,1,2$). The probability to find the system in state $k$ will be denoted by $P_k$.
Transitions between these three states occur due to quasiparticle tunneling. Since charge-$e/3$ quasiparticles tunnel without interference, the corresponding tunneling rate, to be denoted by $\Gamma_1$, is independent of either the enclosed flux or the accumulated charge at the drain D$_2$. This is not the case for charge-$2e/3$ quasiparticles, for which the tunneling rate corresponding to a transition $k \to k+2$ (mod 3) may be written as
$\Gamma_2^k(\Phi) = \Gamma_{2} + \Gamma_{2}^\prime \cos(4\pi \Phi_{AB}/(3 \Phi_0) - 4\pi k/3 + \phi_2)$, with positive $\Gamma_{2}^\prime \le \Gamma_{2}$, where equality is reached only for $|L_d-L_u|/v_c \ll \max[1/T, 1/|eV|]$.
The ensuing master equation is then:
\begin{equation}
  \frac{\mathrm{d}}{\mathrm{d}t}
  \left(
  \begin{matrix}
    p_0 \\ p_1 \\ p_2
  \end{matrix}
  \right)
  =
  \left(
  \begin{matrix}
    -\Gamma_1 -\Gamma_2^0 & \Gamma_2^1 & \Gamma_1 \\
    \Gamma_1 & -\Gamma_1 -\Gamma_2^1 & \Gamma_2^2 \\
    \Gamma_2^0 & \Gamma_1 & -\Gamma_1 -\Gamma_2^2
  \end{matrix}
  \right)
  \left(
  \begin{matrix}
     p_0 \\ p_1 \\ p_2
  \end{matrix}
  \right).
\end{equation}
The steady state current $I_{D2} = |e|/3 \sum_k (\Gamma_1 + 2 \Gamma_2^k) P_k$ is
\begin{widetext}
\begin{equation}
  I_{D2}
  =
  \frac{|e|}{3} \left\{ \Gamma_1 + 2 
  \frac{
  \left( \Gamma_1 \right)^2 \Gamma_{2}
  + \Gamma_1 \left[ \left( \Gamma_{2} \right)^2 - \left( \frac{\Gamma_{2}^\prime}{2} \right)^2 \right]
  + \left[ \left( \Gamma_{2} \right)^3 - 3 \Gamma_{2} \left( \frac{\Gamma_{2}^\prime}{2} \right)^2 + 2 \left( \frac{\Gamma_{2}^\prime}{2} \right)^3 \cos \left(2 \pi \Phi_{AB}/\Phi_0 + 3 \phi_2 \right) \right]
  }{
  \left( \Gamma_1 \right)^2
  + \Gamma_1 \Gamma_{2}
  + \left[ \left( \Gamma_{2} \right)^2 - \left( \frac{\Gamma_{2}^\prime}{2} \right)^2 \right]
  } \right\}.
\end{equation}
\end{widetext}
Hence, if the tunneling amplitude of charge-$2e/3$ quasiparticles is smaller than the corresponding amplitude for charge-$e/3$ quasiparticles (as one physically expects), the visibility (ratio of the amplitudes of the flux-dependent and flux independent current) scales as $(\Gamma_{2}^\prime/\Gamma_1)^3/2$, that is, as the \emph{sixth} power of the tunneling amplitudes ratio, which could be quite small in practice \modified{(it is of order $10^{-3}$ already for a 1/3 amplitude ratio)}.

\emph{Conclusion.---}
The specific analysis presented here concerns a minimal model for the edge of the $\nu=2/3$ FQH state, that supports a charged downstream moving mode as well as a neutral upstream mode. The tunneling of quasiparticles within a MZI setup is accompanied by the emission of upstream neutral excitations.  This leads to the suppression (exponential in the temperature $T$, power-law in the voltage $V$; cf.\ Table~\ref{tbl:visibility} and Fig.~\ref{fig:visibility}) of the AB interference signal.
To facilitate the observation of anyonic interferometry one thus needs to resort to small interferometers at low values of $T$ and $V$, namely
$L_d/v_c + \max(L_d,L_u)/v_n \ll \hbar/(k_B T), \hbar/|eV|$.
\modified{The main unknowns here are the velocities: the only available experimental data is for the integer QH 
regime, with a charge velocity in the $10^6$~cm/sec range \cite{neder07b}. Due to the smaller gap, FQH charge modes are expected  to be slower; the velocities of the softer neutral modes should be smaller still. Taking a typical MZI size of $5$~$\mu$m together with $v_n \approx 10^5$~cm/sec, the corresponding voltage and temperature scales are 0.1~$\mu$V and $1.5$~mK, respectively. Hence, neutral modes should indeed present a significant challenge for current generation MZIs. We are aware, however, of present experimental efforts to reduce the MZI size down to 
the micron scale \cite{heiblum}. This would bring the required temperature and voltage to the presently accessible regime.}

The reason for the adverse role played by these neutral excitations is that they serve as ``markers'' for specific interference paths, much in the spirit of ``which path'' detection. This interference suppression mechanism is thus of general validity: It applies to  more complex (multi-mode) edge structure \cite{wang13}, a host of other fractional states featuring upstream 
neutral modes \cite{inoue14}, setups supporting nonabelian excitations \cite{dinaii15}, and scenarios with charged upstream modes \modified{(e.g., clean filling-2/3 edges)}.
Allowing for tunneling operators which do not excite the upstream modes (specifically in our model, charge-$2e/3$ tunneling) will still lead to suppression (as a high power of the corresponding tunneling amplitude ratio) of the interference signal due to ``dressing'' by neutral-excitation emitting quasiparticles.
Upstream neutral modes are ubiquitous in the FQH regime \cite{inoue14}. They therefore lead to a ``universal'' suppression mechanism of interference signals in the MZ geometry for FQH states \cite{fn:fb_diff}.

\emph{Acknowledgments.---}
We would like to thank L.I. Glazman, M. Heiblum, Y. Meir, and H.-S. Sim for useful discussions. M. G. acknowledges support from the Israel Science Foundation (ISF, Grant No. 227/15) and the German-Israeli Foundation (GIF, Grant No. I-1259-303.10/2014). Y. G. has been supported by the DFG (RO 2247/8-1 and CRC 183), ISF (Grant No. 1349/14), and the Minerva Foundation. Both Authors further acknowledge support by the Israel-Russia Collaboration Program of the Israel Ministry of Science and Technology (MOST) under Contract No. 3-12419.


%




\renewcommand{\thesection}{S.\Alph{section}}
\setcounter{figure}{0}
\renewcommand{\thefigure}{S\arabic{figure}}
\setcounter{equation}{0}
\renewcommand{\theequation}{S\arabic{equation}}

\begin{widetext}

\section*{Supplementary Material for ``Suppression of interference in quantum Hall Mach-Zehnder geometry by upstream neutral modes''}


\modified{%
In this Supplemental Material we present some technical aspects which were omitted in the main text.
Sec.~\ref{sec:visibility_gen} contains a detailed calculation of the visibility in the general case. The limit $\min(L_d,L_u)(1/v_c+1/v_n) \gg 1/T$ is treated in Sec.~\ref{sec:visibility_lowt}, while the case of equal interferometer arms, $L_u=L_d=L$ is addressed in Sec.~\ref{sec:visibility_bal}; this leads to the results presented in
Table.~II
and
Fig.~2
of the main text.
Sec.~\ref{sec:visibility_discuss} presents a qualitative discussion of some general features of the results.
Finally, analysis of the almost closed QPC limit (as opposed to the nearly open QPC limit discussed in the main text) is described in Sec.~\ref{sec:electron}.%
}

\modified{%
\section{Calculation of the visibility: The general case} \label{sec:visibility_gen}}

Here we present a detailed calculation of the visibility, based on 
Eqs.~(3)--(6)
of the main text. This requires the evaluation of the time integral in 
Eq.~(5),
which we will do in the complex $t$ plane.
We consider general, non-integral values of $\kappa_c$ and $\kappa_n$ (the values relevant for $\nu=2/3$ are listed in
Table I,
where the singularities of $F(y_d,y_u,t;T)$ appearing in 
Eqs.~(5)--(6)
are branch cuts with branch points at $y_d/v_c + i m/T + i \delta_{d,c}$, $y_u/v_c + i m/T + i\delta_{u,c}$, $-y_d/v_n + i m/T + i \delta_{d,n}$, and $-y_u/v_n + i m/T + i\delta_{u,n}$, where $m \in \mathbb{Z}$ and $\delta_{d/u,c/n}$ are positive infinitesimals.
The second line of 
Eq.~(6)
defines $F(y_d,y_u,t;T)$ on the real $t$-axis \{See e.g., Appendix~H of Ref.~\cite{delft98}, for an explicit calculation. The formulas appearing there require three modifications for our use: (i) the imaginary time $\tau$ is replaced by $i t$; (ii) $x$ is replaced by $-y_d/v_c$, $-y_u/v_c$, $y_d/v_n$, or $y_u/v_n$, respectively; (iii) $\sigma \equiv \mathrm{sgn}(\tau)$ should be omitted, since we are interested in the greater Green function $G^>$, rather than in the time ordered one\}.
The branch cuts of $F(y_d,y_u,t;T)$ in the complex plane must be chosen in a way consistent with this value of $F(y_d,y_u,t;T)$ on the real $t$ axis but are otherwise arbitrary.

For $F(0,0,t;T)$ (i.e., $y_d=y_u=0$) the four sets of branch points merge into one set, at $i m/T + i \delta$. It is then useful to choose the branch cuts of all the fractional-power functions appearing $F(y_d,y_u,t;T)$ (namely, $z ^{\kappa_c^2}$ and \modified{$z ^{\kappa_n^2}$}) to be along the negative imaginary $z$ axis, with the argument of $z$ defined as $-\pi/2$ and $3\pi/2$ on the right and left sides of the cut, respectively.

We employ the rectangular contour depicted in Fig.~\ref{fig:contours}(a), 
which includes a ``tongue'' going around the $m=0$ branch cut. Let us assume that $\kappa_c^2,\kappa_n^2$ are small enough so that the small circle around the branch point do not contribute; the final result can be analytically continued to the general case.
With the above choices, the part of the contour along the line $\text{Im}(t)=1/T$ will give $-e^{i 2 \pi (\kappa_c^2 + \kappa_n^2)}e^{-e^* V/T}$ times the part of the contour along the real $t$ axis.
We then get:
\begin{align} \label{eqn:j00_general}
  j_{\kappa_c^2,\kappa_n^2}(0,0;V,T) & =
  - 2 i e^* e^{i\pi(\kappa_c^2 + \kappa_n^2)} \sin[2\pi (\kappa_c^2 + \kappa_n^2)]
  (\pi T \tau)^{2(\kappa_c^2 + \kappa_n^2)}
  \frac{1 - e^{- e^* V / T} }
  {1 - e^{i 2 \pi (\kappa_c^2 + \kappa_n^2) - e^* V/T}}
  \int_{0}^\infty \text{d}t e^{i e^* V t }
  \frac{ 1 } {\sinh^{2(\kappa_c^2 + \kappa_n^2)} ( \pi T  t ) }
  \nonumber \\
  & =
  \frac{ e^* \sinh[e^* V/(2T)] \sin[2\pi (\kappa_c^2 + \kappa_n^2)]}
  {\sin \left[ \pi(\kappa_c^2 + \kappa_n^2) + i \frac{e^* V}{2T} \right]}
  \frac{(2 \pi T \tau)^{2(\kappa_c^2 + \kappa_n^2)}}{\pi T}
  B \left( 1 - 2 \kappa_c^2 - 2 \kappa_n^2, \kappa_c^2 + \kappa_n^2 - i \frac{e^* V}{2 \pi T} \right)
  \nonumber \\
  & =
  2 e^* \tau \sinh[e^* V/(2T)]
  (2 \pi T \tau)^{2(\kappa_c^2 + \kappa_n^2)-1}
  \frac{\left| \Gamma \left( \kappa_c^2 + \kappa_n^2 + i \frac{e^* V}{2 \pi T}\right) \right|^2}{\Gamma \left[ 2(\kappa_c^2 + \kappa_n^2) \right]},
\end{align}
where $B(x,y)$ and $\Gamma(z)$ are the Beta and Gamma functions, respectively \cite{Sgradshteyn}.
Thus, for $|e^* V| \ll T$ we have 
\begin{equation} \label{eqn:j00_T}
  j_{\kappa_c^2,\kappa_n^2}(0,0;V,T) =
  2 \pi \tau^2 (e^*)^2 V
  (2 \pi T \tau)^{2(\kappa_c^2 + \kappa_n^2 - 1 )}
  \frac{\left[ \Gamma ( \kappa_c^2 + \kappa_n^2 ) \right]^2}{\Gamma \left( 2\kappa_c^2 + 2\kappa_n^2 \right) }
  \sim
  V T^{2(\kappa_c^2 + \kappa_n^2 - 1 )},
\end{equation}
while for $|e^* V| \gg T$, remembering that $\Gamma(z+a)/\Gamma(z+b) \sim z^{a-b}$ for large $|z|$, we find
\begin{equation} \label{eqn:j00_V}
  j_{\kappa_c^2,\kappa_n^2}(0,0;V,T) =
  \frac{(e^*)^2 V \tau^2}{\Gamma(2\kappa_c^2 + 2\kappa_n^2)}
  \left( |e^* V| \tau \right)^{2(\kappa_c^2 + \kappa_n^2 - 1)}
  \sim V^{2(\kappa_c^2 + \kappa_n^2) - 1 }.
\end{equation}

\begin{figure}
\includegraphics[width=\textwidth,height=!]{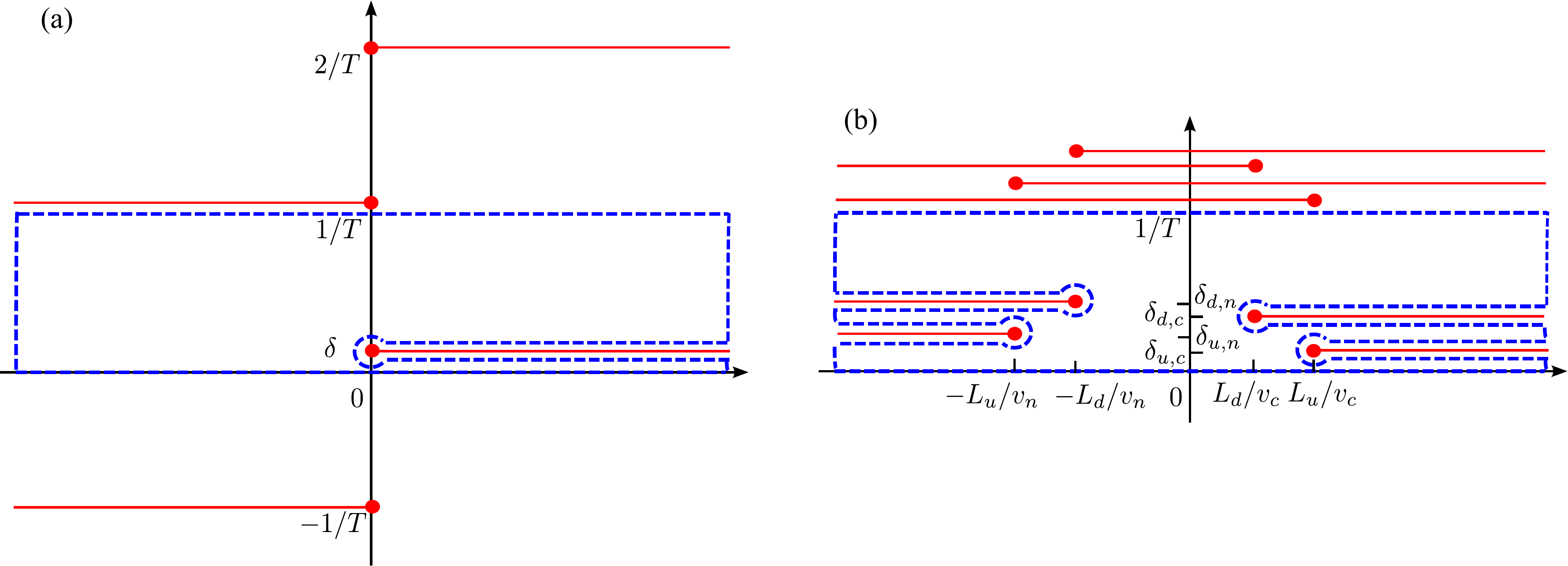}
\caption{\label{fig:contours}
Contours of integration in the complex time plane for the calculation of:
(a) $j_{\kappa_c^2,\kappa_n^2}(0,0;V,T)$, Eq.~(\ref{eqn:j00_general});
(b) $j_{\kappa_c^2,\kappa_n^2}(L_d,L_u;V,T)$, Eq.~(\ref{eqn:j_general}).
Branch points and lines are depicted in red, the contours (traversed in a counterclockwise fashion) in dashed-blue.
The infinitesimal shifts $\delta$ and $\delta_{u/d,c/n}$ of the imaginary parts of the branch points from integer multiples of $1/T$ are exaggerated for clarity.
See the text for further details.
}
\end{figure}

For $j(L_d,L_u,t;T)$ (i.e., $y_d=L_d$ and $y_u=L_u$)
it will be useful to choose 
the branch cut of the fractional power function $z ^{\kappa_c^2}$ as before, that is to be along the negative imaginary axis, with the argument of $z$ defined as $-\pi/2$ and $3\pi/2$ on the right and left sides of the cut, respectively.
On the other hand, the branch cut of the fractional power function $z^{\kappa_n^2}$ will be chosen along the positive imaginary axis, with the argument of $z$ defined as $\pi/2$ or $-3\pi/2$ on the right and left sides of the cut, respectively.
With this the function $F(L_d,L_u,t;T)$ will have the following branch structure [cf.~Fig.~\ref{fig:contours}(b)]: The branch cuts starting at the branch points at $L_d/v_c + i m/T + i \delta_{d,c}$ and $L_u/v_c + i m/T + i\delta_{u,c}$ will go horizontally to the right (left) for even (odd) $m$, and the branch cuts staring at the branch points at $-L_d/v_n + i m/T + i \delta_{d,n}$ and $-L_u/v_n + i m/T + i\delta_{u,n}$ to go horizontally to the left (right) for even (odd) $m$. We will focus on $L_u>L_d$, and take for definiteness $\delta_{d,c} > \delta_{u,c}$ and $\delta_{d,n} > \delta_{u,n}$.
For the right- (left-)going branches the phase of the argument of the corresponding fractional power is $\mp \pi/2$ below the cut and $\pm 3\pi/2$ above it.
Since $\kappa_c^2,\kappa_n^2 < 1$ 
(cf.~Table~I),
the small circles around the branch points do not contribute (by analytical continuation, the final results can be shown to be valid even without this restriction).
With the above choices, the part of he contour along the line $\text{Im}(t)=1/T$ will give $- e^{i 2 \pi (\kappa_c^2 - \kappa_n^2)}e^{-e^* V/T}$ times the part of the contour along the real $t$ axis.
We thus find
\begin{align} \label{eqn:j_general}
  j_{\kappa_c^2,\kappa_n^2}(L_d,L_u; & V,T) =
  - e^*
  \frac{1-e^{- e^* V / T}}{1 - e^{i 2 \pi (\kappa_c^2 - \kappa_n^2) - e^* V/T}}
  \int_{-\infty}^\infty \text{d}t e^{i e^* V (t-L_d/v_c) }
  \times
  \nonumber \\ &
  \left\{
  e^{-i\pi\kappa_n^2}
  \left(e^{i 3 \pi \kappa_c^2/2} - e^{-i \pi \kappa_c^2/2 }\right)
  \left[\theta(t-L_d/v_c) \theta(L_u/v_c-t) e^{i \pi \kappa_c^2/2}
  + \theta(t-L_u/v_c) \left( e^{i 3 \pi \kappa_c^2/2} + e^{- i \pi \kappa_c^2/2} \right)
  \right]
  \right. \nonumber \\ & \left.
  +e^{i\pi\kappa_c^2}
  \left(e^{-i 3 \pi \kappa_n^2/2} - e^{i \pi \kappa_n^2/2 }\right)
  \left[
  \theta(t+L_u/v_n) \theta(-L_d/v_n-t) e^{-i \pi \kappa_n^2/2}
  + \theta(-L_u/v_n-t) \left( e^{-i 3 \pi \kappa_n^2/2} + e^{i \pi \kappa_n^2/2} \right)
  \right]
  \right\}
  \times \nonumber \\ &
  \left\{
  \frac{ (\pi T \tau)^2}{\sinh \left[ \pi T | t - L_d/v_c | \right]
  \sinh \left[ \pi T | t - L_u/v_c | \right]}
  \right\}^{\kappa_c^2}
  \left\{
  \frac{ (\pi T \tau)^2 }{\sinh \left[ \pi T | t + L_d/v_n | \right]
  \sinh \left[ \pi T | t + L_u/v_n | \right]}
  \right\}^{\kappa_n^2}.
\end{align}
This general expression can be simplified in particular cases, as detailed in the following two Sections.

\modified{%
\section{Visibility of a long-arms interferometer}
\label{sec:visibility_lowt}
An explicit expression for Eq.~(\ref{eqn:j_general}) can be obtained for
$\min(L_d,L_u)(1/v_c+1/v_n) \gg 1/T$, allowing for arbitrary $|L_u-L_d|$.}
In this region, on the cuts corresponding to the neutral propagators the  hyperbolic sines appearing in the charge propagators may be replaced by exponentials and vice versa. The integral thus reduces to:
\begin{align}
  j_{\kappa_c^2,\kappa_n^2} & (L_d,L_u; V,T) =
  -2 i e^* e^{i\pi(\kappa_c^2-\kappa_n^2)}
  (2 \pi T \tau)^{2(\kappa_c^2 + \kappa_n^2)}
  \frac{1-e^{- e^* V / T}}
  {1 - e^{i 2 \pi (\kappa_c^2 - \kappa_n^2) - e^* V/T}}
  \int_{-\infty}^\infty \text{d}t e^{i e^* V (t-L_d/v_c) }
  \times \\ &
  \left\{
      \vphantom{ \frac{ e^{-\pi \kappa_n^2 T (2t + L_d/v_n + L_u/v_n)} }
      {\left\{ 4 \sinh \left[ \pi T | t - L_d/v_c | \right] \sinh \left[ \pi T | t - L_u/v_c | \right] \right\}^{\kappa_c^2} } }
  \sin ( \pi \kappa_c^2 )
  \left[ \theta(t-L_d/v_c) \theta(L_u/v_c-t)
  + 2 \theta(t-L_u/v_c) \cos( \pi \kappa_c^2 ) \right]
  \frac{ e^{-\pi \kappa_n^2 T (2t + L_d/v_n + L_u/v_n)} }
  {\left\{ 4 \sinh \left[ \pi T | t - L_d/v_c | \right]
    \sinh \left[ \pi T | t - L_u/v_c | \right] \right\}^{\kappa_c^2} }
  \right. \nonumber \\ & \left. \nonumber
  - \sin (\pi \kappa_n^2)
  \left[
  \theta(t+L_u/v_n) \theta(-L_d/v_n-t)
  + 2 \theta(-L_u/v_n-t) \cos(\pi \kappa_n^2)
  \right]
  \frac{ e^{\pi \kappa_c^2 T (2t - L_d/v_c - L_u/v_c)} }
  {\left\{ 4 \sinh \left[ \pi T | t + L_d/v_n | \right]
  \sinh \left[ \pi T | t + L_u/v_n | \right] \right\}^{\kappa_n^2} }
  \right\}.
\end{align}

Each of the resulting integrals (on the intervals $t<-L_u/v_n$, $-L_u/v_n < t < -L_d/v_n$, $L_d/v_c<t<L_u/v_c$, and $L_u/v_n<t$, respectively) can be calculated exactly and expressed in terms of the hypergeometric function ${}_2F_1(a, b, c; z)$ \cite{Sgradshteyn} through an appropriate change of variables and the Euler integral:
\begin{equation}
  \int_0^1 w^{b-1} (1-w)^{c-b-1} (1-z w)^{-a} \text{d}w
  = B(b,c-b) {}_2F_1(a,b,c;z).
\end{equation}
Employing the relations
\begin{align}
  {}_2F_1(a,b,c;z) = & \frac{\Gamma(c) \Gamma(c-a-b)} {\Gamma(c-a) \Gamma(c-b)} {}_2F_1(a,b,a+b-c+1,1-z) +
  (1-z)^{c-a-b} \frac{\Gamma(c) \Gamma(c-a-b)} {\Gamma(c-a) \Gamma(c-b)} {}_2F_1(a,b,c-a-b+1,1-z),
  \\
  {}_2F_1(a,b,c;z) = & (1-z)^{c-a-b} {}_2F_1(c-a, c-b, c; z),
\end{align}
to bring the argument $z$ of all the hypergeometric functions to the same form, the result can be recast as:
\begin{align} \label{eqn:j_hyper}
  j_{\kappa_c^2,\kappa_n^2} & (L_d,L_u;V,T) =
  -\frac{i e^* e^{i\pi(\kappa_c^2 - \kappa_n^2)}}{\pi T} (2 \pi T \tau)^{2\kappa_c^2} (2 \pi T \tau)^{2\kappa_n^2}
  \Gamma(1-\kappa_c^2)
  \frac{1-e^{- e^* V /T}}
  {1 - e^{i 2 \pi (\kappa_c^2 - \kappa_n^2) - e^* V/T}}
  e^{-i e^* V  L_d/v_c}
  \times \nonumber \\ &
  \left\{
  \left[ \sin(2 \pi \kappa_c^2)
  \frac{\Gamma \left( \kappa_c^2 + \kappa_n^2 - i \frac{e^* V}{2\pi T} \right)}
  {\Gamma \left(1 + \kappa_n^2 - i \frac{e^* V}{2 \pi T} \right)}
  \right.
  + \sin(\pi \kappa_c^2)
  \frac{\Gamma \left( -\kappa_n^2 + i\frac{e^* V}{2\pi T}\right)}
  {\Gamma \left( 1 - \kappa_c^2 - \kappa_n^2 + i \frac{e^* V}{2\pi T} \right)}
  \right]
  \times \nonumber \\ &
  e^{i e^* V L_u/v_c
  -\pi \kappa_n^2 T (2L_u/v_c + L_d/v_n + L_u/v_n)
  - \pi \kappa_c^2 T (L_u-L_d)/v_c}
  {}_2F_1 \left( \kappa_c^2, \kappa_c^2 + \kappa_n^2 - i \frac{e^* V}{2\pi T}, 1 + \kappa_n^2 - i\frac{e^* V}{2\pi T} ; e^{-2 \pi T (L_u-L_d)/v_c} \right)
  \nonumber \\ &
  +
  \left[
  \sin(\pi \kappa_c^2)
  \frac{\Gamma \left( \kappa_n^2 - i \frac{e^* V}{2\pi T}\right)}
  {\Gamma \left( 1-\kappa_c^2+\kappa_n^2 - i \frac{e^* V}{2\pi T} \right)}
  \right]
  \times \nonumber \\ &
  e^{ i e^* V L_d/v_c
  -\pi \kappa_n^2 T (2L_d/v_c + L_d/v_n + L_u/v_n)
  -\pi \kappa_c^2 T (L_u-L_d)/v_c}
  {}_2F_1 \left( \kappa_c^2, \kappa_c^2 - \kappa_n^2 + i \frac{e^* V}{2\pi T}, 1 - \kappa_n^2 + i\frac{e^* V}{2\pi T} ; e^{-2 \pi T (L_u-L_d)/v_c} \right)
  \nonumber \\ & \left.
  - \left[ c \leftrightarrow n, V \to -V \right]
     \vphantom{\frac{\Gamma \left( \kappa_c^2 + \kappa_n^2 - i \frac{e^* V}{2\pi T} \right)}
     {\Gamma \left(1 + \kappa_n^2 - i \frac{e^* V}{2 \pi T} \right)}}
  \right\}
\end{align}
rearranging and using the identity $\Gamma(z)\Gamma(1-z) = \pi/\sin(\pi z)$ \cite{Sgradshteyn} we find
\begin{align} \label{eqn:j_lxy_gen}
  j_{\kappa_c^2,\kappa_n^2} & (L_d,L_u;V,T) =
  -\frac{e^*}{2 T} (2 \pi T \tau)^{2\kappa_c^2} (2 \pi T \tau)^{2\kappa_n^2}
  e^{e^* V/(2 T) - i e^* V L_d/v_c} 
  \left[ 1-e^{- e^* V /T} \right]
  \times \nonumber \\ &
  \left\{
  \frac{1}{\sin \left( \pi \kappa_n^2 - i \frac{e^* V}{2 T} \right)} \frac{\Gamma \left( \kappa_c^2 + \kappa_n^2 - i \frac{e^* V}{2\pi T} \right)}
  {\Gamma(\kappa_c^2) \Gamma \left(1 + \kappa_n^2 - i \frac{e^* V}{2 \pi T} \right)}
  \right.
  \times \nonumber \\ &
  e^{ie^* V L_u/v_c
  -\pi \kappa_n^2 T (2L_u/v_c + L_d/v_n + L_u/v_n)
  - \pi \kappa_c^2 T (L_u-L_d)/v_c}
  {}_2F_1 \left( \kappa_c^2, \kappa_c^2 + \kappa_n^2 - i \frac{e^* V}{2\pi T}, 1 + \kappa_n^2 - i \frac{e^* V}{2\pi T} ; e^{-2 \pi T (L_u-L_d)/v_c} \right)
  \nonumber \\ &
  -
  \frac{1}{\sin \left( \pi \kappa_n^2 - i \frac{e^* V}{2 T} \right)}
  \frac{\Gamma \left( \kappa_c^2 - \kappa_n^2 + i \frac{e^* V}{2\pi T}\right)}
  {\Gamma(\kappa_c^2) \Gamma \left( 1-\kappa_n^2 + i \frac{e^* V}{2\pi T} \right)}
  \times \nonumber \\ &
  e^{ ie^* V L_d/v_c
  -\pi \kappa_n^2 T (2L_d/v_c + L_d/v_n + L_u/v_n)
  -\pi \kappa_c^2 T (L_u-L_d)/v_c}
  {}_2F_1 \left( \kappa_c^2, \kappa_c^2 - \kappa_n^2 + i \frac{e^* V}{2\pi T}, 1 - \kappa_n^2 + i\frac{e^* V}{2\pi T} ; e^{-2 \pi T (L_u-L_d)/v_c} \right)
  \nonumber \\ & \left.
  - \left[ n \leftrightarrow c, V \to -V \right]
       \vphantom{\frac{\Gamma \left( \kappa_c^2 + \kappa_n^2 + i e^* \frac{V}{2\pi T} \right)}
       {\Gamma \left(1 + \kappa_n^2 + i \frac{e^* V}{2 \pi T} \right)}}
  \right\}.
\end{align}

In the limit $L_u-L_d \gg \max(v_c,v_n)/T $ the hypergeometric functions approach unity. We are then left with
\begin{align}
  j_{\kappa_c^2,\kappa_n^2} & (L_d,L_u;V,T) =
  -\frac{e^*}{2 T} (2 \pi T \tau)^{2\kappa_c^2} (2 \pi T \tau)^{2\kappa_n^2}
  e^{e^* V/(2 T) - i e^* V L_d/v_c} 
  \left[ 1-e^{- e^* V / T} \right]
  \times \nonumber \\ &
  \left\{
  \frac{1}{\sin \left( \pi \kappa_n^2 - i \frac{e^* V}{2 T} \right)} \frac{\Gamma \left( \kappa_c^2 + \kappa_n^2 - i \frac{e^* V}{2\pi T} \right)}
  {\Gamma(\kappa_c^2) \Gamma \left(1 + \kappa_n^2 - i \frac{e^* V}{2 \pi T} \right)}
  \right.
  e^{i e^* V L_u/v_c
  -\pi \kappa_n^2 T (2L_u/v_c + L_d/v_n + L_u/v_n)
  - \pi \kappa_c^2 T (L_u-L_d)/v_c}
  \nonumber \\ &
  -
  \frac{1}{\sin \left( \pi \kappa_n^2 - i \frac{e^* V}{2 T} \right)}
  \frac{\Gamma \left( \kappa_c^2 - \kappa_n^2 + i \frac{e^* V}{2\pi T}\right)}
  {\Gamma(\kappa_c^2) \Gamma \left( 1-\kappa_n^2 + i \frac{e^* V}{2\pi T} \right)}
  e^{ i e^* V L_d/v_c
  -\pi \kappa_n^2 T (2L_d/v_c + L_d/v_n + L_u/v_n)
  -\pi \kappa_c^2 T (L_u-L_d)/v_c}
  \nonumber \\ & \left.
  - \left[ c \leftrightarrow n, V \to -V \right]
       \vphantom{\frac{\Gamma \left( \kappa_c^2 + \kappa_n^2 + i \frac{e^* V}{2\pi T} \right)}
       {\Gamma \left(1 + \kappa_n^2 + i \frac{e^* V}{2 \pi T} \right)}}
  \right\}.
\end{align}
In this case, for $|e^* V| \ll T$ we obtain
\begin{align}
  j_{\kappa_c^2,\kappa_n^2}(L_d,L_u;V,T) = &
  -e^* V \frac{(2 \pi \tau)^{2 (\kappa_n^2 + \kappa_c^2)} }{2 \pi} T^{2(\kappa_c^2 + \kappa_n^2 -1)} 
  e^{- ie^* V L_d/v_c} 
  \times \nonumber \\ &
  \left\{
  B(-\kappa_n^2, \kappa_c^2 + \kappa_n^2 )
  e^{ie^* V L_u/v_c
  -\pi \kappa_n^2 T (2L_u/v_c + L_d/v_n + L_u/v_n)
  - \pi \kappa_c^2 T (L_u-L_d)/v_c}
  \right.
  \nonumber \\ &
  +
  B(\kappa_n^2, \kappa_c^2 - \kappa_n^2 )
  e^{ ie^* V L_d/v_c
  -\pi \kappa_n^2 T (2L_d/v_c + L_d/v_n + L_u/v_n)
  -\pi \kappa_c^2 T (L_u-L_d)/v_c}
  \nonumber \\ & \left.
  - \left[ c \leftrightarrow n, V \to -V \right]
  \right\},
\end{align}
whereas for $|e^* V| \gg T$ \{recalling that $\Gamma(a+z)/\Gamma(b+z) \sim z^{a-b}$ for large $z$ \cite{Sgradshteyn}\} we find:
\begin{align}
   j_{\kappa_c^2,\kappa_n^2}(L_d,L_u;V,T) = &
  -\frac{i e^*}{T} (2 \pi T \tau)^{2\kappa_c^2} (2 \pi T \tau)^{2\kappa_n^2}
  e^{ - ie^* V L_d/v_c} 
  \times \nonumber \\ &
  \left\{
  \frac{ e^{-i \pi \kappa_n^2 \text{sgn}(e^* V)} }{\Gamma(\kappa_c^2)}
  \left( - i e^* \frac{V}{2 \pi T} \right)^{\kappa_c^2 - 1}
  e^{i e^* V L_u/v_c
  -\pi \kappa_n^2 T (2L_u/v_c + L_d/v_n + L_u/v_n)
  - \pi \kappa_c^2 T (L_u-L_d)/v_c}
  \right.
  \nonumber \\ &
  -
  \frac{e^{-i \pi \kappa_n^2 \text{sgn}(e^* V)}}{\Gamma(\kappa_c^2)}
  \left( i e^* \frac{V}{2 \pi T} \right)^{\kappa_c^2 - 1}
  e^{ i e^* V L_d/v_c
  -\pi \kappa_n^2 T (2L_d/v_c + L_d/v_n + L_u/v_n)
  -\pi \kappa_c^2 T (L_u-L_d)/v_c}
  \nonumber \\ & \left.
  - \left[ c \leftrightarrow n, V \to -V \right]
       \vphantom{\frac{\Gamma \left( \kappa_c^2 + \kappa_n^2 + i \frac{e^* V}{2\pi T} \right)}
       {\Gamma \left(1 + \kappa_n^2 + i \frac{e^* V}{2 \pi T} \right)}}
  \right\}.
\end{align}

Returning to Eq.~(\ref{eqn:j_lxy_gen}), in the limit
$L_u - L_d \ll \min(v_c, v_n) / \max(T,|e^* V|)$, we may use
\begin{equation}
  {}_2F_1(a,b,c;1) =
  \frac{\Gamma(c) \Gamma(c-a-b)} {\Gamma(c-a) \Gamma(c-b)},
\end{equation}
to obtain:
\begin{align}
  j_{\kappa_c^2,\kappa_n^2}(L_d,L_u;V,T) = &
  -\frac{e^*}{2 \pi T} (2 \pi T \tau)^{2\kappa_c^2} (2 \pi T \tau)^{2\kappa_n^2}
  e^{e^* V/(2 T) - i e^* V L_d/v_c} 
  \left[ 1 - e^{- e^* V / T} \right]
  \times \nonumber \\ &
  \left\{
  \frac{\Gamma(1 - 2 \kappa_c^2) \sin (\pi \kappa_c^2)}{\sin \left( \pi \kappa_n^2 - i \frac{e^* V}{2 T} \right)}
  \frac{\Gamma \left( \kappa_c^2 + \kappa_n^2 - i \frac{e^* V}{2\pi T} \right)}
  {\Gamma \left(1 -\kappa_c^2 + \kappa_n^2 - i \frac{e^* V}{2 \pi T} \right)}
  \right.
  e^{ie^* V L_u/v_c
  -\pi \kappa_n^2 T (2L_u/v_c + L_d/v_n + L_u/v_n)}
  \nonumber \\ &
  -
  \frac{\Gamma(1 - 2 \kappa_c^2) \sin(\pi \kappa_c^2) }{\sin \left( \pi \kappa_n^2 - i \frac{e^* V}{2 T} \right)}
  \frac{\Gamma \left( \kappa_c^2 - \kappa_n^2 + i \frac{e^* V}{2\pi T}\right)}
  {\Gamma \left( 1 - \kappa_c^2 - \kappa_n^2 + i \frac{e^* V}{2\pi T} \right)}
  e^{ i e^* V L_d/v_c
  -\pi \kappa_n^2 T (2L_d/v_c + L_d/v_n + L_u/v_n)}
  \nonumber \\ & \left.
  - \left[ c \leftrightarrow n, V \to -V \right]
       \vphantom{\frac{\Gamma \left( \kappa_c^2 + \kappa_n^2 + i \frac{e^* V}{2\pi T} \right)}
       {\Gamma \left(1 + \kappa_n^2 + i \frac{e^* V}{2 \pi T} \right)}}
  \right\}.
\end{align}
In the limit $|e^* V| \ll T$ we now have
\begin{align}
  j_{\kappa_c^2,\kappa_n^2}(L_d,L_u;V,T) = &
  -(e^*)^2 V \frac{\Gamma(1 - 2 \kappa_c^2)}{2 \pi } T^{2(\kappa_c^2 + \kappa_n^2 - 1)}
  e^{- i e^* V L_d/v_c} 
  \times \nonumber \\ &
  \left\{
  \frac{\sin (\pi \kappa_c^2)}{\sin (\pi \kappa_n^2)}
  B (1 - 2 \kappa_c^2, \kappa_c^2 + \kappa_n^2 )
  \right.
  e^{i e^* V L_u/v_c
  -\pi \kappa_n^2 T (2L_u/v_c + L_d/v_n + L_u/v_n)}
  \nonumber \\ &
  -
  \frac{ \sin(\pi \kappa_c^2) }{\sin ( \pi \kappa_n^2)}
  B ( 1 - 2\kappa_c^2, \kappa_c^2 - \kappa_n^2 )
  e^{ i e^* V L_d/v_c
  -\pi \kappa_n^2 T (2L_d/v_c + L_d/v_n + L_u/v_n)}
  \nonumber \\ & \left.
  - \left[ c \leftrightarrow n, V \to -V \right]
       \vphantom{\frac{\Gamma \left( \kappa_c^2 + \kappa_n^2 + i \frac{e^* V}{2\pi T} \right)}
       {\Gamma \left(1 + \kappa_n^2 + i \frac{e^* V}{2 \pi T} \right)}}
  \right\},
\end{align}
while for $|e^* V| \gg T$ we get
\begin{align}
  j_{\kappa_c^2,\kappa_n^2} (L_d,L_u;V,T) = &
  -\frac{i e^*}{2 \pi T} (2 \pi T \tau)^{2\kappa_c^2} (2 \pi T \tau)^{2\kappa_n^2}
  e^{- i e^* V L_d/v_c} 
  \times \nonumber \\ &
  \left\{
  \Gamma(1 - 2 \kappa_c^2) \sin (\pi \kappa_c^2) e^{-i \pi \kappa_n^2 \text{sgn}(e^* V)}
  \left( - i \frac{e^* V}{2\pi T} \right)^{2\kappa_c^2 - 1}
  \right.
  e^{i e^* V L_u/v_c
  -\pi \kappa_n^2 T (2L_u/v_c + L_d/v_n + L_u/v_n)}
  \nonumber \\ &
  -
  \Gamma(1 - 2 \kappa_c^2) \sin(\pi \kappa_c^2) e^{-i \pi \kappa_n^2 \text{sgn}(e^* V)}
  \left(i e^* \frac{V}{2\pi T} \right)^{2\kappa_c^2 - 1}
  e^{ i e^* V L_d/v_c
  -\pi \kappa_n^2 T (2L_d/v_c + L_d/v_n + L_u/v_n)}
  \nonumber \\ & \left.
  - \left[ \rho \leftrightarrow \sigma, V \to -V \right]
       \vphantom{\frac{\Gamma \left( \kappa_c^2 + \kappa_n^2 + i \frac{e^* V}{2\pi T} \right)}
       {\Gamma \left(1 + \kappa_n^2 + i \frac{e^* V}{2 \pi T} \right)}}
  \right\}.
\end{align}

If, on the other hand, $\max(v_c,v_n)/|e^* V| \ll (L_u-L_d) \ll \min(v_c, v_n)/T$, we return to Eq.~(\ref{eqn:j_lxy_gen}).
We first use the relation $\lim_{w \to \infty} {}_2F_{1}(a,b_0+w,c_0+w;z) = (1-z)^{-a}$, then expand in $T(L_u-L_d)/v_{c/n}$ to find
\begin{align} \label{eqn:j_oscillate}
  j_{\kappa_c^2,\kappa_n^2}(L_d,L_u;V,T) = &
  -\frac{i e^*}{T} (2 \pi T \tau)^{2\kappa_c^2} (2 \pi T \tau)^{2\kappa_n^2}
  e^{- i e^* V L_d/v_c} 
  \times \nonumber \\ &
  \left\{
  e^{-i \pi \kappa_n^2 \text{sgn}(e^* V)}
  \left(- i \frac{e^* V}{2\pi T}\right)^{\kappa_c^2-1}
  \frac{1}{\Gamma(\kappa_c^2)}
  \right.
  e^{i e^* V L_u/v_c
  -\pi \kappa_n^2 T (2L_u/v_c + L_d/v_n + L_u/v_n)}
  \left(2 \pi T (L_u-L_d)/v_c \right)^{-\kappa_c^2}
  \nonumber \\ &
  -
  e^{-i \pi \kappa_n^2 \text{sgn}(e^* V)}
  \frac{1}{\Gamma(\kappa_c^2)}
  \left( + i \frac{e^* V}{2\pi T}\right)^{\kappa_c^2-1}
  e^{ i e^* V L_d/v_c
  -\pi \kappa_n^2 T (2L_d/v_c + L_d/v_n + L_u/v_n)}
  \left(2 \pi T (L_u-L_d)/v_c \right)^{-\kappa_c^2}
  \nonumber \\ & \left.
  - \left[ c \leftrightarrow n, V \to -V \right]
       \vphantom{\frac{\Gamma \left( \kappa_c^2 + \kappa_n^2 + i \frac{e^* V}{2\pi T} \right)}
       {\Gamma \left(1 + \kappa_n^2 + i \frac{e^* V}{2 \pi T} \right)}}
  \right\}.
\end{align}

\modified{%
\section{Visibility of an equal-arms interferometer}
\label{sec:visibility_bal}
Here we examine in greater detail the case of equal-arms interferometer, $L_d=L_u=L$.}
We allow for arbitrary relation between $L/v_c + L/v_n$, $1/T$, and $1/(e^* V)$. Returning to Eq.~(\ref{eqn:j_general}), it now becomes
\begin{align}
  & j_{\kappa_c^2,\kappa_n^2}(L,L;V,T) =
  -2i e^* e^{i \pi (\kappa_c^2-\kappa_n^2)}
  \frac{1 - e^{- e^* V / T} }
  {1 - e^{i 2 \pi (\kappa_c^2 - \kappa_n^2) - e^* V/T}}
  \int_{-\infty}^\infty \text{d}t e^{i e^* V (t-L/v_c) }
  \times
  \nonumber \\ &
  \left\{
   \sin (2 \pi \kappa_c^2) \theta(t-L/v_c)
  - \sin (2 \pi \kappa_n^2) \theta(-L/v_n - t)
  \right\}
  \left\{
  \frac{ \pi T \tau }{\sinh \left[ \pi T | t - L/v_c | \right]}
  \right\}^{2 \kappa_c^2}
  \left\{
  \frac{ \pi T \tau }{\sinh \left[ \pi T | t + L/v_n | \right]}
  \right\}^{2\kappa_n^2}.
\end{align}
Employing the same methods as before, the integral over each of the regimes $t>L/v_c$ and $t<-L/v_n$ can be expressed exactly in terms of hypergeometric functions, resulting in:
\begin{align} \label{eqn:j_bal_hyper}
  j_{\kappa_c^2,\kappa_n^2} (L,L;V,T) = &
  2 \pi i e^* \tau \left( 2 \pi T \tau \right)^{2(\kappa_c^2 + \kappa_n^2)-1}
  \times
  \nonumber \\ &
  \left\{
  e^{- 2 \pi \kappa_n^2 T (L/v_c + L/v_n) }
  \frac{\sinh \left( \pi \frac{e^* V}{2 \pi T} \right) }{\sinh\left( \pi \frac{e^* V}{2 \pi T} - i \pi (\kappa_c^2-\kappa_n^2) \right) }
  \frac{\Gamma \left( \kappa_c^2 + \kappa_n^2 - i \frac{e^* V}{2\pi T} \right) }
  {\Gamma(2 \kappa_c^2) \Gamma \left( 1 - \kappa_c^2 + \kappa_n^2 - i \frac{e^* V}{2\pi T} \right)}
  \times
  \right. \nonumber \\ &
  {}_2F_1 \left( 2 \kappa_n^2, \kappa_c^2 + \kappa_n^2 - i \frac{e^* V}{2\pi T}, 1 - \kappa_c^2 + \kappa_n^2 - i \frac{e^* V}{2\pi T}; e^{-2\pi T (L/v_c + L/v_n) } \right)
  \nonumber \\ & 
  -e^{- i e^* V (L/v_c + L/v_n) - 2 \pi \kappa_c^2 T (L/v_c + L/v_n)}
  \frac{\sinh \left( \pi \frac{e^* V}{2 \pi T} \right) }{\sinh\left( \pi \frac{e^* V}{2 \pi T} - i \pi (\kappa_c^2-\kappa_n^2) \right) }
  \frac{\Gamma \left( \kappa_c^2 + \kappa_n^2 + i \frac{e^* V}{2\pi T} \right) }
  {\Gamma(2 \kappa_n^2) \Gamma \left( 1 + \kappa_c^2 - \kappa_n^2 + i \frac{e^* V}{2\pi T} \right)}
  \times
  \nonumber \\ &
  \left.
  {}_2F_1 \left( 2 \kappa_c^2, \kappa_c^2 + \kappa_n^2 + i \frac{e^* V}{2\pi T}, 1 + \kappa_c^2 - \kappa_n^2 + i \frac{e^* V}{2\pi T}; e^{-2\pi T (L/v_c + L/v_n) } \right)
  \right\}.
\end{align}

For $L/v_c + L/v_n \gg 1/T$ we then find
\begin{align}
  j_{\kappa_c^2,\kappa_n^2}(L,L;V,T) = &
  2 \pi i e^* \tau \left( 2 \pi T \tau \right)^{2(\kappa_c^2 + \kappa_n^2)-1}
  \times
  \nonumber \\ &
  \left\{
  e^{- 2 \pi \kappa_n^2 T (L/v_c + L/v_n) }
  \frac{\sinh \left( \pi \frac{e^* V}{2 \pi T} \right) }{\sinh\left( \pi \frac{e^* V}{2 \pi T} - i \pi (\kappa_c^2-\kappa_n^2) \right) }
  \frac{\Gamma \left( \kappa_c^2 + \kappa_n^2 - i \frac{e^* V}{2\pi T} \right) }
  {\Gamma(2 \kappa_c^2) \Gamma \left( 1 - \kappa_c^2 + \kappa_n^2 - i \frac{e^* V}{2\pi T} \right)}
  \right.
  \nonumber \\ & 
  \left.
  -e^{ - i e^* V (L/v_c + L/v_n) - 2 \pi \kappa_c^2 T (L/v_c + L/v_n)}
  \frac{\sinh \left( \pi \frac{e^* V}{2 \pi T} \right) }{\sinh\left( \pi \frac{e^* V}{2 \pi T} - i \pi (\kappa_c^2-\kappa_n^2) \right) }
  \frac{\Gamma \left( \kappa_c^2 + \kappa_n^2 + i \frac{e^* V}{2\pi T} \right) }
  {\Gamma(2 \kappa_n^2) \Gamma \left( 1 + \kappa_c^2 - \kappa_n^2 + i \frac{e^* V}{2\pi T} \right)}
  \right\}.
\end{align}
In this case, in the limit $T \gg |e^* V|$ one obtains
\begin{align}
  & j_{\kappa_c^2,\kappa_n^2}(L,L;V,T) =
  - 2 \pi (e^*)^2 V \tau \left( 2 \pi T \tau \right)^{2(\kappa_c^2 + \kappa_n^2 - 1)}
  \times
  \nonumber \\ &
  \left\{
  e^{- 2 \pi \kappa_n^2 T (L/v_c + L/v_n) }
  B(\kappa_n^2+\kappa_c^2,\kappa_n^2-\kappa_c^2)
  +e^{- i e^* V (L/v_c + L/v_n) - 2 \pi \kappa_c^2 T (L/v_c + L/v_n)}
  B(\kappa_c^2+\kappa_n^2,\kappa_c^2-\kappa_n^2)  \right\},
\end{align}
and in the limit $T \ll |e^* V|$ we have
\begin{align}
  & j_{\kappa_c^2,\kappa_n^2}(L,L;V,T) =
  2 \pi i e^* \tau \left( 2 \pi T \tau \right)^{2(\kappa_c^2 + \kappa_n^2)-1}  e^{i \pi (\kappa_c^2-\kappa_n^2) \text{sgn}(e^* V)}
  \times
  \nonumber \\ &
  \left\{
  e^{- 2 \pi \kappa_n^2 T (L/v_c + L/v_n) }
  \frac{1}{\Gamma(2 \kappa_c^2)}
  \left( - i \frac{e^* V}{2\pi T} \right)^{2 \kappa_c^2 - 1}
  -e^{- i e^* V (L/v_c + L/v_n) - 2 \pi \kappa_c^2 T (L/v_c + L/v_n)}
  \frac{1}{\Gamma(2 \kappa_n^2)}
  \left( i \frac{e^* V}{2\pi T} \right)^{2\kappa_n^2-1}
  \right\}.
\end{align}.

For $L/v_c + L/v_n \ll 1/T, 1/|e^* V|$ we find
\begin{align}
  j_{\kappa_c^2,\kappa_n^2}(L,L;V,T) =
  2 i e^* \tau \left( 2 \pi T \tau \right)^{2(\kappa_c^2 + \kappa_n^2)-1}
  & \left\{
  \frac{\sin(2 \pi \kappa_c^2)\sinh \left( \pi \frac{e^* V}{2 \pi T} \right) }{\sinh\left( \pi \frac{e^* V}{2 \pi T} - i \pi (\kappa_c^2-\kappa_n^2) \right) }
  \frac{\Gamma(1 - 2 \kappa_c^2 - 2 \kappa_n^2) \Gamma \left( \kappa_c^2 + \kappa_n^2 - i \frac{e^* V}{2\pi T} \right) }
  { \Gamma \left( 1 - \kappa_c^2 - \kappa_n^2 - i \frac{e^* V}{2\pi T} \right)}
  \right.
  \nonumber \\ & 
  \left.
  -\frac{\sin(2 \pi \kappa_n^2) \sinh \left( \pi \frac{e^* V}{2 \pi T} \right) }{\sinh\left( \pi \frac{e^* V}{2 \pi T} - i \pi (\kappa_c^2-\kappa_n^2) \right) }
  \frac{\Gamma(1 - 2 \kappa_c^2 - 2 \kappa_n^2) \Gamma \left( \kappa_c^2 + \kappa_n^2 + i \frac{e^* V}{2\pi T} \right) }
  { \Gamma \left( 1 - \kappa_c^2 - \kappa_n^2 + i \frac{e^* V}{2\pi T} \right)}
  \right\},
\end{align}
which can be shown \{using the identity $\Gamma(z)\Gamma(1-z) = \pi/\sin(\pi z)$ \cite{Sgradshteyn}\} to reduce to Eqs.~(\ref{eqn:j00_general})--(\ref{eqn:j00_V}), as it should.

Finally, for $1/|e^* V| \ll L/v_c + L/v_n \ll 1/T$ we have:
\begin{align} \label{eqn:j_bal_oscillate}
  & j_{\kappa_c^2,\kappa_n^2}(L,L;V,T) =
  2 \pi i e^* \tau \left( 2 \pi T \tau \right)^{2(\kappa_c^2 + \kappa_n^2)-1} e^{i \pi (\kappa_c^2-\kappa_n^2) \text{sgn}(e^* V)}
  \times
  \nonumber \\ &
  \left\{
  \frac{1}{\Gamma(2 \kappa_c^2)}
  \left( - i \frac{e^* V}{2\pi T} \right)^{2\kappa_c^2-1}
  \left[ 2\pi T (L/v_c + L/v_n) \right]^{-2\kappa_n^2}
  -e^{- i e^* V (L/v_c + L/v_n)}
  \frac{1}{\Gamma(2 \kappa_n^2)}
  \left( i \frac{e^* V}{2\pi T} \right)^{2\kappa_n^2-1}
  \left[ 2\pi T (L/v_c + L/v_n) \right]^{-2\kappa_c^2}
  \right\}.
\end{align}

From the above results, we can determine the visibility of the interference pattern as function of temperature and source-drain bias, using 
Eqs. (3)--(4).
For $|\gamma_a|=|\gamma_b|$ it is the ratio $|j(L_d,L_u;V,T)/j(0,0;V,T)|^3$,
reflecting the fact that a unit charge detected at D$_2$ 
[Fig.~1(a)]
involves the tunneling of three $e/3$ quasi-particles, each subject to a different effective AB flux. The results are displayed in 
Table.~II
and
Fig.~2
of the main text.

\modified{%
\section{Some qualitative features of the visibility}
\label{sec:visibility_discuss}
There are some remarkable insights that follow from the results of the previous Sections.
First and foremost,}
the interference signal is exponentially-suppressed with increasing temperature as soon as $\min(L_d,L_u)(1/v_c+1/v_n) \gg 1/T$. This is true even even if $T \ll |e^* V|$ (!). At the same time, increasing the bias voltage does not lead to exponential suppression even when $|e^* V|$ is larger than any other scale, including the temperature.
This can be intuitively understood by thinking about free fermions with velocity $v$: at zero temperature the voltage defines a window of allowed energies, and hence wavevectors, for tunneling. Since the window is sharp, the corresponding wavepackets decay in real space at large distances $x$ only as $e^{i V x/v}/x$, hence there is no exponential suppression of the interference. On the other hand, introducing a finite temperature amounts to smearing the sharp energy window. Roughly, this can be thought of as having a distribution of voltage values around some average $V_0$. Averaging $e^{iV x/v}$ with respect to this distribution will leave us with an oscillatory factor $e^{i V_0 x/v}$ multiplied by a suppression factor. The latter is a function of $T x/v$, and decays exponentially when $T x/v \gg 1$, independently of the value of $V_0$.

\modified{%
Another noteworthy feature is that the two terms in Eq.~(\ref{eqn:j_general}), and correspondingly in Eqs.~(\ref{eqn:j_hyper}) and~(\ref{eqn:j_bal_hyper}), originate from domains in the time integration where either the charge or the neutral mode overlap is maximal, but not both (which is the essence of the dephasing mechanism discussed in this work), cf.\ Fig.~\ref{fig:contours}(b). These two terms have a relative phase between them. This gives rise to oscillations of the visibility as function of the voltage when the latter is high and the temperature small [so that the the two terms in Eqs.~(\ref{eqn:j_hyper}) and~(\ref{eqn:j_bal_hyper}) are comparable], the case described by Eqs.~(\ref{eqn:j_oscillate}) and~(\ref{eqn:j_bal_oscillate}). This leads to the last line of
Table.~II
and the lobe structure seen in
Fig.~2
of the main text.%
}

\begin{figure}
\includegraphics[trim=0cm 0cm 0cm 0cm,width=10cm,height=!]{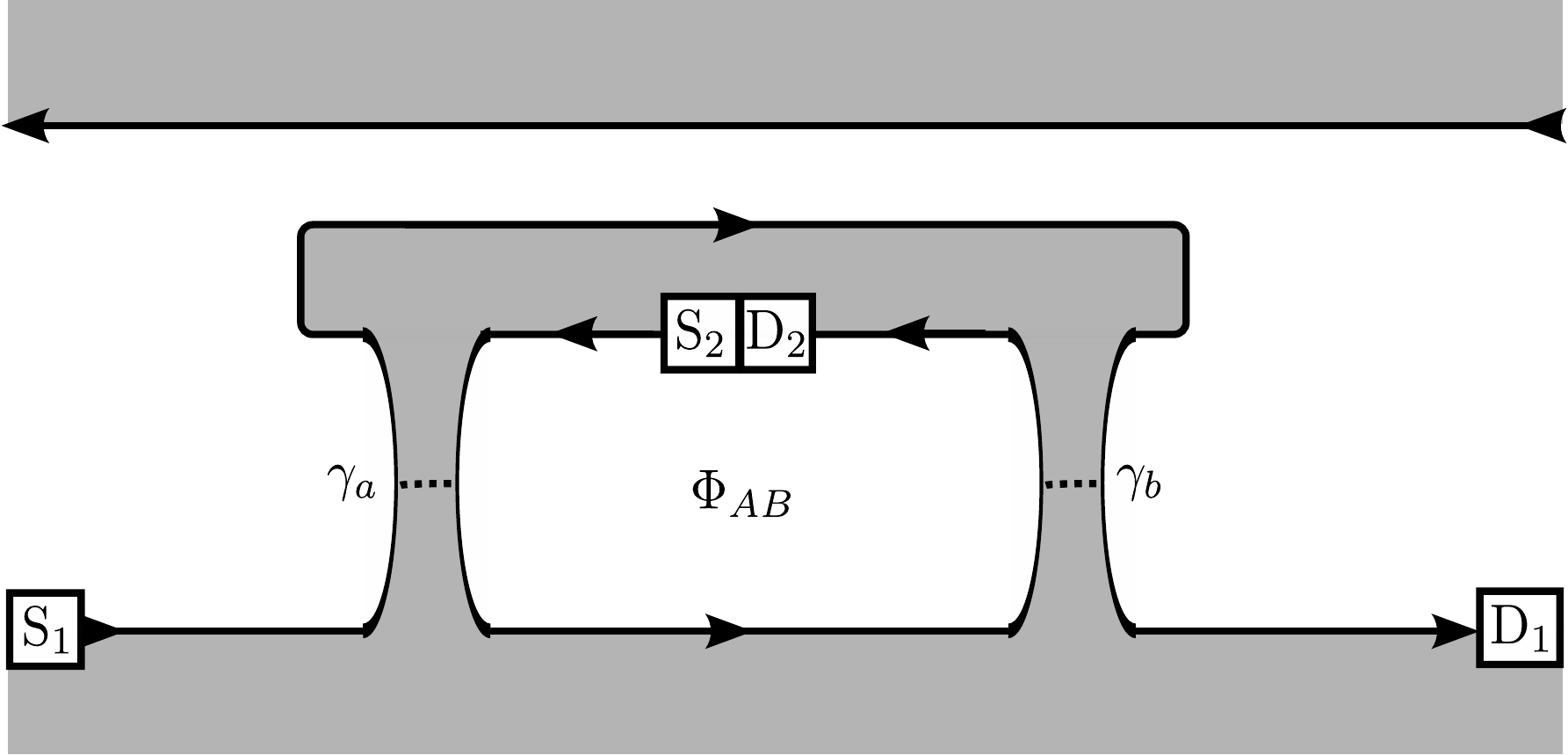}
\vspace{30pt}
\caption{\label{fig:interference_electron}
An almost-closed QH MZI (compare with the almost-open case, 
Fig.~1 of the main text).
Shown are the downstream chiral edge modes, sources, drains, and the tunneling bridges. The schematic equivalent geometry is essentially the same as the one depicted in 
Fig.~1(b)--(c),
with the upper and lower edges corresponding, respectively, to the inner and outer edges of the lower edge + island depicted here, instead of the island and lower edges in 
Fig.~1(a).
}
\end{figure}

\section{The case of electron tunneling} \label{sec:electron}

In the limit of strong tunneling bridges, the interfering paths are dominated by electron tunneling. It is easy to obtain results for the correlation functions by noting that $L_d$ and $L_u$ are interchanged and the AB phase is flipped as compared with quasiparticle tunneling (cf.~Fig.~\ref{fig:interference_electron}).
In addition, the values of $\kappa_c$ and $\kappa_n$ are now taken from the fourth, rather than the first row of 
Table~I.
Finally, 
Eqs. (3)--(4)
are replaced by
\begin{equation} \label{eqn:ie}
  I_{D2}(V,T) =
  \left(\left| \gamma_a \right|^2 + \left| \gamma_b \right|^2 \right) j(0,0;V,T)
  + \gamma_a \gamma_b^* e^{-2 \pi i \Phi_{AB}/\Phi_0 } j(L_u,L_d;V,T)
  + \text{c.c.}
\end{equation}
Thus, when $|\gamma_a|=|\gamma_b|$ the visibility is $|j(L_u,L_d;V,T)/j(0,0;V,T)|$. The quantities $j(L_u,L_d;V,T)$ and $j(0,0;V,T)$ were calculated in Secs.~\ref{sec:visibility_gen}--\ref{sec:visibility_bal}. The resulting behavior is summarized in
Table.~II
of the main text.

\end{widetext}


\begin{thebibliography}{99}
\bibitem{fn:upstream}
In the latter case the signal was about an order of magnitude weaker.
\bibitem{fn:disorder}
The spatially-dependent gauge transformation used to eliminate the disorder potential \cite{kane94,kane95} mixes up the quasiparticle operators at one of the QPCs (choosing the location of the other as the origin). However, to lowest order in the tunneling its only effect is an order-unity reduction of the bare amplitude of the dominant 
tunneling operator. 
\bibitem{fn:three}
The fact that a unit charge detected at D$_2$ [Fig.1(a)] involves the tunneling of three $e/3$ quasi-particles, each subject to a different effective AB flux \cite{law06}, causes the visibility to scale as $|j(L_d,L_u;V,T)/j(0,0;V,T)|^3$.
\bibitem{fn:fb_diff}
The path marker mechanism discussed here is distinctly different from the physics of interference suppression in Fabry-P\'{e}rot interferometers \cite{bishara08,yang08,park15}.
\end{thebibliography}
\end{document}